\documentclass[11pt]{article}
\usepackage{graphicx}
\usepackage{longtable,lscape}
\usepackage{amsthm}
\usepackage{amsfonts}
\usepackage{amsmath}
\usepackage{bbm}
\usepackage{float}
\usepackage{url}

\newcommand{\captionfonts}{\footnotesize}
\makeatletter  % Allow the use of @ in command names
\long\def\@makecaption#1#2{%
  \vskip\abovecaptionskip
  \sbox\@tempboxa{{\captionfonts #1: #2}}%
  \ifdim \wd\@tempboxa >\hsize
    {\captionfonts #1: #2\par}
  \else
    \hbox to\hsize{\hfil\box\@tempboxa\hfil}%
  \fi
  \vskip\belowcaptionskip}
\makeatother 
\begin{document}
\title{What is Quantum? Unifying Its Micro-Physical \\ and Structural Appearance}
\author{Diederik Aerts$^1$ and Sandro Sozzo$^{1,2}$ \vspace{0.5 cm} \\ 
        \normalsize\itshape
        $^1$ Center Leo Apostel for Interdisciplinary Studies \\
        \normalsize\itshape
        and, Department of Mathematics, Brussels Free University \\ 
        \normalsize\itshape
         Krijgskundestraat 33, 1160 Brussels, Belgium \\
        \normalsize
        E-Mail: \url{diraerts@vub.ac.be} \\
        \normalsize\itshape
        $^2$ School of Management, University of Leicester \\
        \normalsize\itshape
         University Road LE1 7RH, Leicester, United Kingdom \\
        \normalsize
        E-Mail: \url{ss831@le.ac.uk}
           \\
              }
\date{}
\maketitle
\begin{abstract}
\noindent
We can recognize two modes in which `quantum appears' in macro domains: (i) a {\it micro-physical appearance}, where quantum laws are assumed to be universal and they are transferred from the micro to the macro level if suitable {\it quantum coherence} conditions (e.g., very low temperatures) are realized, (ii) a {\it structural appearance}, where no hypothesis is made on the validity of quantum laws at a micro level, while genuine quantum aspects are detected at a structural-modeling level. In this paper, we inquire into the connections between the two 
appearances. We put forward the explanatory hypothesis that, 
`the appearance of quantum in both cases' is due to `the existence of a specific form of organisation, which has the capacity to cope with random perturbations that would destroy this organisation when not coped with'.
We analyse how `organisation of matter', `organisation of life', and `organisation of culture', play this role each in their specific domain of application, point out the importance of evolution in this respect, and put forward how our analysis sheds new light on `what quantum is'.
\end{abstract}

\section{Introduction\label{intro}}
\vspace{-2mm}
The strange quantum world unveils every day more its mysterious aspects to us. On one hand, increasing evidence confirms that, whenever entities on large scales are pushed in delicate and specific ways to show quantum effects, such as entanglement, nonlocality, interference, and Bose or Fermi identity, they reveal aspects of this quantum behavior \cite{kapitza1938,london1938,bardeen1957,rauch1975,aspect1982,andersonetal1995,tittel1998,arndt1999,salart2008,gerlich2011,plumhof2014}. Such experiments have reached the astonishing scales of distances of 18 kilometers in the case of entanglement, sizes of large macro- and bio-molecules in the case of interference \cite{salart2008,gerlich2011}, and room temperature realisations of Bose-Einstein condensates \cite{plumhof2014}. On the other hand different aspects of the structure of quantum theory are identified, its probability model, but also the structure of interference and entanglement, and the Bose and Fermi behavior of identity, in situations with entities that are part of the macroscopical world surrounding us. More specifically in human cognition and human decision processes, and in cultural entities such as languages, but also in situations in biology, economics and computer science, 
such typical quantum structures have been found \cite{aertsaerts1995,vanrijsbergen2004,aertsczachor2004,aertsgabora2005a,aertsgabora2005b,widdows2006,melucci2008,aerts2009a,aerts2009b,bruzaetal2009,khrennikov2009,aertsdhooghehaven2010,piwowarski2010,songetal2011,as2011,bpft2011,bb2012,ags2012,pb2013,haven2013,abckss2014}.

These two ways in which `quantum appears' are looked upon differently, and even give rise to different thought about `what quantum is'. We will call these two appearances `micro-physical' and `structural', respectively.

The `micro-physical appearance' of quantum is always accompanied by an explanation which links it with `quantum in the micro-world', assuming that quantum laws hold universally in this micro-world. Quantum effects can then be detected also in the macroscopic world if suitable conditions of control are verified, these conditions being of different types. The conditions can range from the construction of an interferometer capable of creating interference on the macro level to the cooling down of a gas of bosonic quantum particles making them join into one quantum state, a so called `Bose-Einstein condensate'. This tenet constitutes the basis of the research which flourishes in many areas, namely, quantum computation and information \cite{nielsen2000}, Bose-Einstein condensation \cite{andersonetal1995}, superconductivity \cite{bardeen1957}, superfluidity \cite{kapitza1938,london1938}, and ever more macroscopic realisations of double slit interference and entanglement \cite{salart2008,gerlich2011}.

The `structural appearance' of quantum is identified by the criterion that the considered situation can be modeled by using a quantum-theoretic formalism, %Sandro. as modeling tool, 
without a necessary connection with the quantum nature of particles at a microscopic level. This approach has recently produced important achievements in the study of cognitive processes, in the domain of concept research \cite{aertsgabora2005a,aertsgabora2005b,aerts2009b,as2011}, human decision making \cite{bpft2011,bb2012}, but also by modeling situations in economics \cite{khrennikov2009,aertsdhooghehaven2010,haven2013}, biology and ecology \cite{abckss2014}, and computer science, 
i.e. for information retrieval and natural language processing \cite{vanrijsbergen2004,aertsczachor2004,widdows2006,melucci2008,piwowarski2010}.

Since our research activity has touched both quantum appearances, we are naturally led to wonder whether and how they can be connected. In the present paper, we try to answer this question. We put forward an explanatory hypothesis which makes 
it possible to understand the two quantum appearances as being manifestations of one underlying specific organisational state of reality. Our hypothesis leads also to a challenging view on `what quantum is'. 

The hypothesis that we put forward, first here in short, and in the following more explored in detail, is the following. ``That `quantum appears' is connected with the presence of a specific type of organisation, with the property of being able to cope with the intrinsic destructive aspects of random influences of change perturbing the organisation''. We will call this organisation a `quantum organisation'. Hence, it is an organisation able to save itself from destruction due to random influences of change. Concretely, and for the two quantum appearances that we have mentioned, the micro-physical appearance and the structural appearance, we think of `organisations of matter', `organisations of life', 
% Dirk 
and `organisations of culture', 
% Dirk and `organisations of languages',
and will explain more in detail in the following how these are good examples illustrating our general explanatory hypothesis.

\section{The micro-physical quantum appearance\label{traditional}}
\vspace{-2mm}
The identification of  
what we have called `micro-physical appearance of quantum'
has been historically associated with wave-particle interpretations, and so was the identification of the emergence of quantum effects in the macroscopic physical world. More concretely, it is the original formula by Louis de Broglie $\lambda=h/p$, where $\lambda$ is the de Broglie wave length of an entity with momentum $p$, and $h=6.62 \cdot 10^{-34} J\cdot s$ is Planck's constant \cite{debroglie1923} which is customarily used -- certainly by experimentalists -- to reason about the micro-physical
appearance of quantum also if this happens in 
% Dirk the 
the macro world. The idea is that quantum behavior within a collection of entities
appears when the de Broglie waves of 
these entities can overlap, i.e. when the wavelengths are bigger than the typical distance between the entities. The mechanism imagined within the wave-particle interpretations is 
that with overlapping de Broglie waves, the waves can vibrate in phase, join together to (more or less) form a single wave. 
For a gas of particles, such a situation can only occur at very low temperatures, since with increasing temperature, heat adds energy and hence momentum to each of the particles, so that their de Broglie wave lengths will become smaller and smaller, to the extent that the waves no longer overlap. We stress that the pure effect of becoming smaller is not what makes quantum behavior disappear. It is the non-globally structured way in which the wavelength decreases that destroys the quantum coherence. Indeed, heat is intrinsically a non-structured random way of adding energy, which is why `it is a process profoundly disturbing the quantum coherence'. 
The particles of the gas, that at low temperatures were united into one macroscopically sized de Broglie quantum wave, start to get disconnected, their de Broglie waves being pushed out of phase as a consequence of the collisions with random packets of heat energy. This means that with rising temperature the gas starts  
to become a collection of separated particles, behaving classically with respect to each other. Considering our explanatory hypothesis, the quantum organisation here is correlated with the temperature of the environment, if this temperature is low enough, the micro-quantum realm is able to cope with the random disturbance of bombarding energy packets. Hence, the appearance of quantum behavior at a macroscopic scale for gases at very low temperatures, and disappearance of this quantum behavior, being substituted by classical behavior if temperature rises, is a good example of what we have called `quantum organisation'.

Let us give a short overview of these macroscopic quantum entities that constitute a micro-physical appearance of quantum. In 1917 Einstein proposed the microscopic description for the quantum-mechanical mechanism of 
the `laser' \cite{einstein1917}. This was definitely the first macroscopic quantum entity, and no cooling is needed here. The reason is that only photons are involved, and the random bombarding of heat packets existing at room temperature also consists of photons. Photons scatter only extremely rarely with other photons, which is the reason that the laser does not suffer under heat \cite{enterria2013}. Next to the laser, Bose-Einstein condensates are the entities that have brought the micro-quantum behavior to the macroscopic level, but they need heavy cooling, since they exists of atoms or molecules in a gas. And atoms or molecules are highly disturbed by bombardment of random packets of energy, which means that only when cooling down the gas, and in this way shielding of the bombardment, what we have called a quantum organisation becomes possible. The experimental realisation of a Bose-Einstein condensate came about after a long exciting history of cooling gases to temperatures close to the absolute zero. The phenomena of superfluidity and superconductivity, both already observed more than a century ago by  Kamerlingh Onnes in Leiden, and later studied in more detail by Kapitsa, Meissner, London, Landau, Ginzburg and others, were only stepwise identified as being caused by the quantum `Bose-Einstein condensation' phenomenon, and lead in 1995 finally to a first conscious and identified realisation of such a condensate \cite{andersonetal1995}.

\section{The structural quantum appearance\label{structural}}
\vspace{-2mm}
The attention for the structural appearance of quantum was originally rooted in the 
investigation of the structure of the theory of quantum physics itself from its axiomatic to its operational aspects \cite{mackey1963,piron1976,aerts1986}. An essential step took place in identifying similar types of structures in situations of entities in the macroscopic world surrounding us, without the appearance of this structure being connected in any way to micro-physical aspects of quantum \cite{aerts1986,aertsetal1993,aertsaertsbroekaertgabora2000}. A new important step took place when structural quantum aspects started to get identified in aspects of human thought, more specifically in how the human mind makes decisions \cite{aertsaerts1995}, 
developing further to the fruitful use of the mathematical formalism of quantum theory in Hilbert space to model complex situations of decision making 
\cite{bb2012,pb2013,k2010}. Parallel a succesful quantum-theoretic modeling was elaborated for how the human mind uses conceptual entities, like in a language
\cite{aertsgabora2005a,aertsgabora2005b,aerts2009b,ags2012,sozzo2014}, and genuine quantum aspects, such as `contextuality', `emergence', `entanglement', `interference', `superposition' were identified as responsible of the observed deviations from classical (fuzzy set) logic and probability theory \cite{hampton1988b}. Quantum modeling approaches have also been employed in information retrieval and natural language processing to integrate and generalize latent semantic analysis methods \cite{vanrijsbergen2004,aertsczachor2004,widdows2006,melucci2008,piwowarski2010}. The domain of research that followed from this has now been called `quantum cognition', it concerns the use of the theoretical framework of quantum theory to model situations in human cognition
 and is now emerging as a flourishing domain of research \cite{aertsaerts1995,aertsgabora2005a,aertsgabora2005b,aerts2009b,bruzaetal2009,songetal2011,as2011,bpft2011,bb2012,ags2012,pb2013,aertsaertsbroekaertgabora2000,k2010,wang2013,sozzo2014}. 

The detection of quantum structures occurs at the level of the modeling of cognitive and decision phenomena, which involves the Hilbert space framework of quantum theory. More explicitly, one describes the situations mentioned above by introducing conceptual entities, their states, measurements and the corresponding probabilities of outcomes, and then represents them by using the standard Hilbert space representation of entities, states, measurements and probabilities of outcomes in quantum theory. This means that such a modeling does not presuppose the validity of quantum laws at a microscopic level. And, further, there is no need to suppose that the structural quantum appearance would be due to the existence of microscopic quantum processes occurring in the human brain, although such an hypothesis is no a priori rejected. Due to the specific situations that have been investigated, it has meanwhile been possible to go deeper in the identification of the structural appearance of `quantum' than just the detection of a Hilbert space structure for a fruitful model. Indeed, mechanisms have been identified that make it possible to put forward operational structural definitions for entanglement, interference and Bose or Fermi identity. 

Let us specify these operational mechanisms. 

With respect to entanglement, we investigated its structural appearance 
when concepts combine to form a new concept. We considered the concepts {\it Animal} and {\it Acts} and their combination {\it The Animal Acts}. Then we measured in an experiment 
the relative frequencies of changes of this combined concept to more concrete states, i.e. exemplars, of it \cite{as2011,ags2012}. One set of four exemplars 
that we considered for the concept combination {\it The Animal Acts} are, {\it The Horse Growls}, {\it The Horse Whinnies}, {\it The Bear Growls}, and {\it The Bear Whinnies}. Of the 81 persons that participated in the experiment, there were 4, hence a fraction
% Dirk
of 0.05, which choose {\it The Horse Growls} as the `their preferred good example of {\it The Animals Acts}', and there were 51, hence a fraction of 0.63, who choose {\it The Horse Whinnies}, 
21, hence a fraction of 0.26, who choose {\it The Bear Growls}, and 5, hence a fraction of 0.06, who choose {\it The Bear Whinnies}. This means that 
the two exemplars {\it The Horse Whinnies} and {\it The Bear Growls} were considered to be the preferred good examples of the concept combination {\it The Animal Acts}, which is what we would expect taken into account the `meaning' of the sentence {\it The Animal Acts}. However, if we asked the same participants in the experiment to elect their `preferred good example of {\it Animal} and of {\it Acts}, as separated concept', resulted that 43 of the 81 choose {\it Horse} and 38 choose {\it Bear}, hence respectively fractions 0.53 and 0.47, for the concept {\it Animal}, while 39 choose {\it Growls} and 42 choose {\it Whinnies}, respectively fractions 0.48 and 0.52, for the concept {\it Acts}. If we consider these fractions as estimates of the probabilities of change or collapse, our experiment shows that the combination {\it The Animal Acts} collapses respectively with probabilities 0.05, 0.63, 0.26 and 0.06, to the more concrete states or exemplars of it, namely {\it The Horse Growls}, {\it The Horse Whinnies}, {\it The Bear Growls}, {\it The Bear Whinnies}, within the human minds of the participants of the experiments. However the concepts apart, {\it Animal} and {\it Acts} collapse respectively with probabilities 0.53 and 0.47 to {\it Horse} or {\it Bear}, and with probabilities 0.48 and 0.52 to {\it Growls} or {\it Whinnies}. If both, the collapse mechanism of the combined concept {\it The Animals Acts} to one of the collapsed states 
and the collapse mechanism of the single concepts {\it Animal} and {\it Acts} to a combination of the collapsed states  
would be the same, we would need the four joint probabilities, 0.05, 0.63, 0.26 and 0.06, to be the products of the single probabilities,  0.53 and 0.47, and 0.48 and 0.52. Let us 
see that this is not the case. We have
`just combining without involving meaning' $\leftrightarrow$ ({\it Horse}, {\it Growls}) $\leftrightarrow$ $0.53 \cdot 0.48=0.25 \not=0.05$ $\leftrightarrow$ ({\it The Horse Growls}) $\leftrightarrow$ `meaningfully combining'. Also,
`just combining without involving meaning' $\leftrightarrow$ ({\it Horse}, {\it Whinnies}) $\leftrightarrow$  $0.53 \cdot 0.52=0.28 \not=0.63$ $\leftrightarrow$ ({\it The Horse Whinnies}) $\leftrightarrow$ `meaningfully combining'. 
%Additionally, we have `just combining without involving meaning' $\leftrightarrow$ ({\it Bear}, {\it Growls}) $\leftrightarrow$ $0.47 \cdot 0.48=0.23 \not=0.26$ $\leftrightarrow$ ({\it The Bear Growls}) $\leftrightarrow$ `meaningfully combining'. And finally, `just combining without involving meaning' $\leftrightarrow$ ({\it Bear}, {\it Whinnies}) $\leftrightarrow$
% and $0.47 \cdot 0.52=0.24\not=0.06$ $\leftrightarrow$ ({\it The Bear Whinnies}) $\leftrightarrow$ `meaningfully combining'.
 The same reasoning can be repeated for ({\it Bear}, {\it Growls}) with respect to ({\it The Bear Growls}), and for ({\it Bear}, {\it Whinnies}) with respect to ({\it The Bear Whinnies}),
and results again in the joint probabilities not being products of the single one.

We understand very well why these joint probabilities are not equal to the products of the combined probabilities: it is because the sentence {\it The Animal Acts} carries `meaning', and the minds of the humans participating in the experiment carry also this meaning, which makes the collapses in their minds to more concrete exemplars be guided by this meaning of the combination, and not just be a combination of the collapses that their minds provoke with the single concepts. 
We have proved \cite{as2011} that the way these joint probabilities deviate from being products of the single probabilities makes them violate Bell's inequalities \cite{bell1964}. 
We do not dwell on this here, but we only mention that such a violation of Bell's inequalities proves that the joint probabilities cannot be products of probabilities related to the single component concepts, and cannot be fit into a classical probability structure, which is what entanglement means when it appears in quantum physics. Hence {\it Animal} and {\it Acts} are entangled through meaning in the combination {\it The Animal Acts}.

We also have understood how the structural appearance of interference takes place. 
We have studied the combination of concepts {\it Fruits} and {\it Vegetables} in the disjunction {\it Fruits or Vegetables}. This time however 
% Dirk we ask 
participants in a %Sandro. n experiment
test are asked to choose amongst exemplars that 
are all concrete states of the three concepts, the two single ones, and the combined one.
Interference  
effects results in this experiment. For example, an exemplar such a {\it Olive}, will be chosen much more often for the combination {\it Fruits or Vegetables} than a `logical disjunction analysis' of the data allows, even if we apply quantum logic. The reason is that next to the disjunction {\it Fruits} or {\it Vegetables}, the combination {\it Fruits or Vegetables} is also a new emergent concepts, that gives special weight to the exemplars for which one can doubt whether they are fruits or whether they are vegetables, and {\it Olive} is such an exemplar. 
It is quite amazing 
% Dirk is 
that this effect is captured in a complete way by interference 
of the type encountered in quantum theory. 
And the complex numbers in quantum theory, which make interference much more powerful as compared to how it appears with waves and real numbers, plays a crucial role in the faithful modeling of the data \cite{aerts2009a,aerts2009b,ags2012}.   

We believe that, in the structural appearance of quantum, even more unique quantum aspects 
manifest, such as `how 
 identical quantum entities behave'.
Indeed, although we have demonstrated above entanglement and interference by means of concepts and how they combine, these effects can also appear structurally at the level  
of physical matter, without the need to consider the cognitive realm where the human mind interacts. We have, e.g., presented examples of entanglement by connected vessels of water \cite{aertsaertsbroekaertgabora2000}, and interference is well 
% Dirk know 
known to take place with physical waves in matter. But, the weird 
way in which identical quantum entities behave, we have 
only structurally found back in how concepts behave within the realm of human cognition \cite{aerts2009a,aerts2010b,aerts2014}, and we have good reasons to believe that it only there appears. Indeed, we have an explanation, although speculative, for why it appears in human cognition structurally in the way identified in \cite{aerts2009a,aerts2010b,aerts2014}. 
Our explanation rests on a theory about the evolution of human concepts, where these come into existence when humans develop the capacity to create states of minds for shared intentions during collaborations \cite{tomasello2005}. Although the identification of objects, and the communication about these objects, which usually is thought to be at the origin of concepts, certainly has played a an important role in the primitive stages of human conceptuality, recent research indicates that `shared intentionality' would be the major aspect giving rise to the specifics of this human conceptuality. Following this research, the crucial difference between human cognition and that of other species would be the ability to participate with others in collaborative activities with shared goals and intentions, where participation in such activities would require a unique motivation to share psychological states with others and unique forms of cognitive representation for doing so.
% Dirk 
This results in a species-unique form of cultural cognition including the use of linguistic symbols, construction of social norms and individual beliefs \cite{tomasello2005}. Hence, `shared intentionality' would be the driving force behind human cognition along this 
scenario, resulting in `a human mind with increasing capacity to create internal states representing such shared intentions'. 
We believe that the conceptual representations resulting from such shared intentions carry within them the paradoxical aspects also to be found in the behavior of identical quantum entities. To explain what we mean, let us imagine eleven of our ancestors to be collaborating in hunting. The colaboration will only be successful in case all eleven are able to create a conceptual representation of the hunting scene which is `identical' on the conceptual level -- it represents the same unique hunting scene -- but of course will be (at least slightly) different for each of the eleven minds -- for example, they all will have a  
different role in the hunting activity. The equivalent in quantum theory are eleven fermionic identical quantum entities, being identical, but when actualised within a piece of matter -- the equivalent of the hunting scene actualised in each of the eleven minds -- will always appear in different states, due to the Pauli exclusion principle. In 
a further stage of development of human language, also the bosonic version of quantum identify appears, 
namely when communicated about `eleven hunting events'. 
Indeed, within the communication itself, hence the exchange of concepts, these concepts are identical and can also be in the same state. It is indeed not necessary for eleven minds to be involved to communicate about eleven hunting events, two minds is enough. This is the way we have analysed the concept `eleven animals' and found it to obey a Bose-Einstein statistics \cite{aerts2009a}. We are investigating actually these identity aspects of human concepts including the data of an experiment on human subjects \cite{asForthcoming}. 

The analysis above and in the first section illustrates that in both situations, the one of micro-physical appearance  
and the one of structural appearance of quantum, this `quantum' is destroyed in case random perturbations are allowed to take place, at least if the perturbations are able to provoke a change in the quantum state of the entities involved. On the contrary, quantum persists in case such perturbations are able to be avoided, which can be by shielding of or in other ways, for example by the nature of the organisation itself.
In Section \ref{coherence} we analyse our explanatory hypothesis in additional detail.

\section{Unifying micro-physical and structural appearance\label{coherence}}
\vspace{-2mm}
Let us mention, to initiate the reasoning we will develop in this section, that a Bose-Einstein condensate has recently been fabricated 
% Dirk 
at room temperature -- lasting for a few picoseconds -- by using a thin non-crystalline polymer film of approximately 35 nanometers thick
% Dirk  at room temperature 
\cite{plumhof2014}. Also important for our analysis is that genuine quantum effects of the micro-physical appearance type have been identified in biology, more specifically a quantum tunneling phenomenon in the process of photo-synthesis. Also the effect discovered in biology occurs at room temperature or, better, at earth crust temperature \cite{saravar2010}. 
Since the size of the random bombardment of energy packets of any entity in our surroundings depends crucially on the temperature both cases mentioned above are again good illustrations for our explanatory hypothesis.
Indeed, it is plausible that a plant, in the processes that enable it to use photo-synthesis,  has managed to be less disturbed by this bombardment of random heat packets of energy due to the mechanism of biological evolution that has played a fundamental role in what the plant is, and how photo-synthesis works. 
And what about the appearance of quantum effect in human laboratories at room temperature?
Human culture is also an evolutionary process, albeit not Darwinian. It has not only managed resistance against the random bombardment of heat energy packets, but also evolved to use this heat energy and make it into non-random energy. Humans' energy-harvesting from heat started with the first steam engine, which literally is the transformation of random energy into structured energy. Does this 
gives rise to quantum structure? 
% Dirk
Not always, and not automatically, but this is certainly the case for the energy used in those laboratories that have produced quantum effect at room temperature. What about the vessels of water and other macroscopic situations we invented to violate Bell's inequalities \cite{aertsaertsbroekaertgabora2000}, and the identification of quantum structure in cognition \cite{aertsaerts1995,aertsgabora2005a,aertsgabora2005b,aerts2009b,as2011,sozzo2014}? Well, the vessels of water and the other entities violating Bell's inequalities are realized within human culture, so that they can be said to have been specially devised to violate Bell's inequalities, albeit not in explicit laboratory situations. In doing so, they make use of all knowledge available to achieve this. As regards the presence of quantum structure in human cognition, we note that human cognition is a product of human culture, and hence profits from the mechanism of cultural evolution to fight the destructive effect of 
random perturbations in case these perturbations invoke changes that are destructive for the cognition. A simple example, we avoid to have too much noise in the environment in case we want to have a conversation with someone.
Hence, not only for plants, but more generally, the capacity of living matter to manage destructive effects of bombardments of random energy packets is the product of evolution. 
For plants and photosynthesis it is the consequence of biological evolution, and it takes place even on a semi-microscopic level.  
For animals, and humans, which materially speaking are made of living matter, but additionally have a nervous system, and brains, the interaction with the environment contains primitive and less primitive aspects of conceptuality. For primitive animals, with primitive nervous systems, these interactions create coordinations and/or competitions or collaborations and hence give rise to situations where entanglement and interference appear on the macroscopic level. One could state that a nervous system is 
an amplifier for quantum from the micro-level to the macro-level, because it allows the entity with the nervous system to develop 
complicated strategies of defence against random perturbations with changes that are destructive for the evolved organisation. 
% Dirk In the case, 
In the case of human beings, this capacity of defence has 
evolved to a very sophisticated level, fully exploring the amplifying effect of the nervous system, and giving rise to 
cultural cognition, with languages and other cultural items as  
manifestations of it. This is in our opinion the essence of 
cultural evolution. These effects manifest in the macroscopic world in the two ways we have discussed in Sections \ref{traditional} and \ref{structural}. 
We can even classify the ability of experimentally controlling random bombardments of heat energy packets  
for the construction of suitable 
experimental situations which allow the
emergence of quantum 
in the macro world, as 
a fruit of cultural evolution. 
Equally so, the appearance of quantum structures in human cognition, decision making and language is a consequence of 
humans being able to
organize, transfer and communicate language in a coherent way, without it being destroyed by `random perturbations'.

Consider the situation where you go to a big garbage belt, like the ones one typically finds in a metropolis, and you collect the words belonging to pieces of texts, newspapers, scrambled books, etc., that you find there, and put them in a huge basket. These words are not connected by meaning, they are completely 
random.  
This situation of a `bag of words' can be modeled by using the known classicalities (set theory, Boolean logic, Kolmogorovian product probabilities). Consider instead the situation where you go to a library. This  
is completely different, because  
meaning is keeping purposefully all the words in the books on their one and unique place, as an exemplar consequence of 
human cultural evolution. We now know that quantum aspects will occur in this case, if one collects experimental data on the words belonging to the books in 
that library. Now, take one of these books and cut it in several pieces of paper, corresponding to single words 
in the book, repeat the operation for all the books and mix together the pieces of paper so obtained. The library has in this way taken the form of a `bag of words' which is very similar to a garbage belt, hence one expects that the situation is classical. To demonstrate this concretely, let us perform the experiment on the conceptual combination {\it The Animal Acts} considered in Section \ref{structural} \cite{as2011,ags2012,as2014IJTP}, and ask a subject to report the first combination among {\it The Horse Growls}, {\it The Bear Whinnies}, etc. that he/she finds at random in the pieces of paper in the library. This situation is obviously classical, the joint probabilities for the different exemplars of the combined concepts {\it The Animal Acts}, will all be neatly product probabilities of the exemplars related to the single concepts {\it Animal} and {\it Acts}, because the `bag of words' only contains single words, and not any meaning is left to connect these single words, which means that Bell's inequalities will not be violated, in this case.

In a garbage belt, the quantum organisation of human culture is destroyed, exactly as in a bombardment of random packets of energy at room temperature, the quantum 
organisation occurring at the microscopic level is destroyed. Analogously, the situation of two persons who talk with each other communicating and exchanging meaning, preserves 
the quantum organisation that is identified within human conceptuality and language.
 
Our  unification of the two ways that `quantum appear' should not make us forget that also still differences exists between the two ways. 
In particular, it seems that the entire technical apparatus of Hilbert space is fully  
represented for its micro-physical appearance
% Dirk 
-- although `separated quantum entities' might cause of problem in this respect \cite{aerts2014} --, while only particular 
aspects of it -- although the major ones -- can be identified for its structural appearance. Of course, this difference is also fundamentally due to the structural appearance being defined as `allowing structure to be identified step by step', which means that the existence of this difference should not be seen as a flaw in the analysis we put forward in the present article. It does mean however that some quantum effects notably present in its micro-physical appearance do not find its counterpart -- at least not till now -- in its structural appearance. We only mention the role played by `spin', and its connection to Bose or Fermi identity behavior for what concerns the micro-physical appearance of quantum. This means that, although we believe that in the present article we reveal a crucial new aspect of `what quantum is' with our unification of its micro-physical appearance and its structural appearance, and our explanatory hypothesis of why this unification is possible, still other aspects of `what quantum is' remain open as challenging questions for future research.

\end{document}